\input{epsf}
\documentclass[11pt]{article}
\def\be{\begin{equation}}
\def\eea{\end{eqnarray}}
\def\bea{\begin{eqnarray}}
\def\ee{\end{equation}}

\author{M.Mohammadi$^{1,2}$ \footnote{majid471702@yahoo.com} , M.H.Naderi$^{3}$ \footnote{mhnaderi2001@yahoo.com} and M.Soltanolkotabi$^{3}$ \footnote{soltan@sci.ui.ac.ir}
\\ $^{1}$ {\small  Islamic Azad University Science and Research Branch, Tehran, Iran}
\\$^{2}$ {\small Department of Physics, Shahreza Islamic Azad University, Shahreza, Isfahan, Iran}
\\$^{3}$ {\small Quantum Optics Group, Department of Physics, University of Isfahan, Isfahan, Iran}}
\title{Influence of a classical homogeneous gravitational field on dissipative dynamics of the Jaynes-Cummings
           model with phase damping }
 
\begin{document}
\maketitle
\begin{abstract}
\noindent In this paper, we study the dissipative dynamics of the Jaynes-Cummings
           model with phase damping in the presence of a classical
           homogeneous gravitational field. The model consists of a moving two-level atom simultaneously
exposed to the gravitational field and a single-mode traveling radiation field in the presence of the phase damping.
 We present a quantum
treatment of the internal and external dynamics of the atom based on an alternative su(2) dynamical algebraic
structure. By making use of the super-operator technique, we obtain the solution of the master equation for the
density operator of the quantum system,  under the Markovian approximation.
Assuming that initially the radiation
field is prepared in a Glauber coherent state and the two-level atom is
in the excited state, we investigate the
influence of gravity on the temporal evolution of collapses and revivals of the atomic
population inversion, atomic dipole squeezing, atomic momentum
diffusion, photon counting statistics and quadrature squeezing of
the radiation field in the presence of phase damping.
\end{abstract}
\\
\section{Introduction}
Over the last forty years many theoretical investigations have been addressed toward the understanding of quantum dynamics
of the interacting atom-field system in a high-Q cavity. The interest toward this research area was mainly spurred by the large
amount of experiments revealing the appearance of intriguing features of quantum radiation-matter interaction [1].
Both theoretical and experimental activities have concentrated on trying to understand simple nontrivial models of quantum
optics involving a single atom, regarded as a few effective energy levels, and one or more rear resonant modes of the quantized
electromagnetic field. The prototype of such systems, proposed by Jaynes and Cummings in 1963, [2] describes a two-level
atom resonantly interacting with a single-mode quantized field. It has proved to be a theoretical laboratory of great
relevance to many topics in atomic physics and quantum optics, as well as in the ion traps [3], cavity QED [4] and quantum
information processing [5]. The Jaynes-Cummings model (JCM) also widely used in condensed matter physics for its relevance
in spintronics [6] which exploits the electron-spin rather than its charge to develop a new generation of electronic
devices [7]. When the rotating wave approximation (RWA) is made, the model becomes exactly solvable and its dynamical
features can be analytically brought to light revealing remarkable properties [8]. The discovery of interesting
aspects of the JCM as well as the developments in cavity QED experiments involving single Rydberg atoms within single-mode
cavities, have stimulated an intense research devoted at highlighting and generalizing the original idea and physical
scenario presented by Jaynes and Cummings.
 In the standard JCM, the interaction between a
constant electric field and a stationary (motionless) two-level
atom is considered. With the development in the technologies of
laser cooling and atom trapping the interaction between a moving
atom and the field has attracted much attention [9-18].\\
\hspace*{00.5 cm} Experimentally, atomic beams
with very low velocities are generated in laser cooling and
atomic interferometry [19]. It is obvious that for atoms moving
with a velocity of a few millimeters or centimeters per second
for a time period of several milliseconds or more, the influence
of Earth's acceleration becomes important and cannot be neglected
[20]. To get a clear picture of what is going to happen it may be useful to refer to the equivalence principle.
It states that the influence of a homogeneous gravitational field on the atom moving in a radiation field can be
simulated by constant acceleration. This
means that the following situation is physically equivalent to
the atom-radiation system exposed to a gravity field: An atom is
at rest or moving with constant velocity relative to an inertial
system. The laboratory with the radiation field attached to it
moves with constant acceleration. The consequence is that the
radiation field reaches the atom with Doppler shifted frequency.
Because of the acceleration this shift changes in time. It acts
as a time-dependent detuning. A semi-classical description of a two-level atom interacting with a
running laser wave in a gravitational field has been studied
[21,22]. However, the semi-classical treatment does not permit us
to study the pure quantum effects occurring in the course of
atom-radiation interaction. Recently, within a quantum
treatment of the internal and external dynamics of the atom, we have presented [23] a theoretical scheme based on
an su(2) dynamical algebraic structure to investigate the
influence of a classical homogeneous gravitational field on the quantum
non-demolition measurement of atomic momentum in the dispersive
JCM. Also, we have investigated [24] quantum statistical properties of the lossless Jaynes-Cummings
           model in the presence of a homogeneous gravitational field. We have found that the non-classical properties
           are suppressed with increase of the
gravitational field influence.\\
\hspace*{00.5 cm} On the other hand, over the last two decades much attention has been focused
on the properties of the dissipative variants of the JCM. The theoretical
efforts have been stimulated by experimental progress in cavity QED. Besides the experimental drive,
there also exists a theoretical motivation to include relevant damping mechanism
to JCM because its dynamics becomes more interesting. A number of authors have treated the JCM with
 dissipation by the use of analytic approximations [25,26] and numerical calculations [27-31]. The
solution in the presence of dissipation is not only of theoretical interest, but also important from a practical
point of view since dissipation would be always present in any experimental realization of the model.
 However, the dissipation treated in the above studies is modeled by coupling to an external reservoir including
 energy dissipation. As is well known, in a dissipative quantum system, the system loses energy by creating
 a bath quantum. In this kind of damping the interaction Hamiltonian between bath and system does not commute with the
 system Hamiltonian. In general this leads to a thermalization of the system with a certain time constant.
 There are, however other kinds of environmental coupling to the system, which do not involve energy exchange.
 In the so-called phase damping [32] the interaction Hamiltonian commutes with that of system and
 in the dynamics only the phase of system state is changed in the course of interaction. Similar to standard energy damping the
 off-diagonal elements of the density matrix in energy basis decay at a given rate. The phase damping can well describe
 some unaccounted decay of coherences in a single-mode micromaser [33]. It has also been shown that phase damping
 seriously reduces the fidelity of the received qubit in quantum computers due to the induced decoherence
 [34]. The phase damping in the JCM with one quantized field mode has been studied [35]. The influence of phase damping
 on non-classical properties of the multi-quanta two-mode JCM has also been studied [36]. It has been found
 that the phase damping suppresses non-classical effects of the cavity field in the JCM. However, all of the foregoing
 studies have been done only under the condition that the influence of the gravitational field is not taken into account.\\
 \hspace*{00.5 cm} In the present contribution our main purpose is to investigate the temporal evolution of
quantum statistical properties of the phase damped JCM in the presence of a classical
homogeneous gravity field. In the JCM,
when the atomic motion is in a propagating light wave, we
consider a two-level atom interacting with the quantized
cavity-field with phase damping in the presence of a homogeneous gravitational
field. By solving analytically the master equation under the Markovian approximation, the evolving reduced density operator
of the system is
found by which the influence of the gravitational field on
 the dynamical behavior of the atom-radiation system in the presence of the phase damping is explored.
 In section 2, we present the master equation for the reduced density operator
of the system under Markovian approximation in terms of a Hamiltonian describing the atom-radiation interaction in
the presence of a gravitational field. This Hamiltonian has been obtained based on an
 su(2) dynamical algebraic structure in the interaction picture. In section 3 we obtain an exact solution of
  the JCM with the phase damping in the presence of a gravitational field, by which we
investigate the dynamical evolution of the system.
 In section 4 we
study the influence of the gravitational
field on both the cavity-field and the atomic properties in the presence of phase damping.
Considering the field to be initially in a coherent state and the
two-level atom in the excited state, we explore the temporal evolution of the
atomic inversion, atomic dipole squeezing, atomic momentum
diffusion, photon
counting statistics and quadrature squeezing of the radiation
field. Finally, we summarize our conclusions in section 5.
\section{ Master Equation for the phase damped JCM in the Presence of a Gravitational Field }
The equation of motion for the density operator of the atom-radiation
 system and reservoir, $\hat{\rho}_{sr}(t)$,
  in the Schr\"{o}dinger picture is given by [37]
\begin{equation}
\frac{\partial\hat{\rho}_{sr}(t)}{\partial t}=-i[\hat{\tilde{H}}_{T},\hat{\rho}_{sr}(t)]  (\hbar=1),
\end{equation}
where
\begin{equation}
\hat{\tilde{H}}_{T}=\hat{H}_{s}+\hat{H}_{r}+\hat{V}_{sr},
\end{equation}
with the Hamiltonian of the reservoir
\begin{equation}
\hat{H}_{r}=\sum_{i}\omega_{i}\hat{b}_{i}^{\dag}\hat{b}_{i},
\end{equation}
and with the Hamiltonian of the interaction between the system and reservoir
\begin{equation}
\hat{V}_{sr}=\hat{H}_{s}\sum_{j=1}^{3}\hat{F}_{j},
\end{equation}
where
\begin{equation}
\hat{F}_{1}=\sum_{i}\kappa_{i}\hat{b}_{i}, \hat{F}_{2}=\sum_{i}\kappa_{i}\hat{b}_{i}^{\dag}, \hat{F}_{3}=\hat{H}_{s}\sum_{i}\frac{\kappa_{i}^{2}}{2\omega_{i}},
\end{equation}
$\hat{b}_{i}$ and $\hat{b}_{i}^{\dag}$ are the boson annihilation and creation operators for the reservoir
 and $\kappa_{i}$ is
the coupling constant.
The Hamiltonian $\hat{H}_{s}$ in (2) for the atom-radiation system in the presence of a classical gravity
 field with the atomic
motion along the position
vector $\hat{\vec{x}}$ and in the RWA is given by ($\hbar=1$)
\begin{eqnarray}
\hat{H}_{s}=&&\frac{\hat{p}^{2}}{2M}-M\vec{g}.\hat{\vec{x}}+\omega_{c}(\hat{a}^{\dag}\hat{a}+\frac{1}{2})+\frac{1}{2}\omega_{eg}\hat{\sigma}_{z}+\nonumber\\
&&\lambda[\exp(-i\vec{q}.\hat{\vec{x}})\hat{a}^{\dag}\hat{\sigma}_{-}+\exp(i\vec{q}.\hat{\vec{x}})\hat{\sigma}_{+}\hat{a}],
\end{eqnarray}
where $\hat{a}$ and $\hat{a}^{\dag}$ denote, respectively, the
annihilation and creation operators of a single-mode traveling
wave with frequency $\omega_{c}$, $\vec{q}$ is the wave vector of
the running wave and $\hat{\sigma}_{\pm}$ denote the raising and
lowering operators of the two-level atom with electronic levels
$|e\rangle, |g\rangle $ and Bohr transition frequency
$\omega_{eg}$. The atom-field coupling is given by the parameter
$\lambda$ and
 $\hat{\vec{p}}$, $\hat{\vec{x}}$
denote, respectively, the momentum and position operators of the
atomic center of mass motion and $g$ is Earth's gravitational
acceleration. It has been shown [23] that based on an su(2)
algebraic structure, as the dynamical symmetry group of the
model, Hamiltonian (6) takes the following form
\begin{eqnarray}
\hat{H}_{s} =&&\frac{\hat{p}^{2}}{2M}-M\vec{g}.\hat{\vec{x}}+
\omega_{c}\hat{K}+\frac{1}{2}\triangle\hat{S_{0}}+\lambda\sqrt{\hat{K}}(\exp(-i\vec{q}.\hat{\vec{x}})\hat{S_{-}} \\ \nonumber +&&\exp(
i\vec{q}.\hat{\vec{x}} )\hat{S}_{+}),
\end{eqnarray}
where
the operators
\begin{equation}
\hat{S_{0}}=\frac{1}{2}(|e \rangle \langle e|-|g \rangle \langle
g|) , \hat{S_{+}}=\hat{a}|e\rangle \langle
g|\frac{1}{\sqrt{\hat{K}}},
\hat{S_{-}}=\frac{1}{\sqrt{\hat{K}}}|g\rangle \langle
e|\hat{a}^{\dag},
\end{equation}
with the following commutation relations
\begin{equation}
[\hat{S_{0}},\hat{S_{\pm}}]=\pm
\hat{S_{\pm}},[\hat{S_{-}},\hat{S_{+}}]=-2\hat{S_{0}},
\end{equation}
are the generators of the su(2) algebra, the operator $\hat{K}=\hat{a}^{\dag}\hat{a}+|e\rangle\langle e| $
is a constant of motion which represents the total number of excitations of the atom-radiation and
 $\Delta=\omega_{eg}-\omega_{c}$ is the detuning parameter.
The corresponding time evolution operator for Hamiltonian (7)
can be expressed as [23]
\begin{equation}
\hat{u}(t)=\exp(iM\vec{g}.\hat{\vec{x}}t)\hat{v}^{\dag}\hat{u}_{e}(t)\hat{v},
\end{equation}
where
\begin{equation}
\hat{v}=\exp(-i\vec{q}.\hat{\vec{x}}\hat{S}_{0}), \hat{u}_{e}=\exp(-i\hat{H}'_{s}t).
\end{equation}
It can be shown that the operator $\hat{u}_{e}(t)$ satisfies an
effective Schr$\ddot{o}$dinger equation governed by an effective
Hamiltonian $\hat{H}'_{s}$, that is
\begin{equation}
i\frac{\partial\hat{ u}_{e}}{\partial
t}=\hat{H}'_{s}\hat{u}_{e},
\end{equation}
where
\begin{eqnarray}
\hat{H}'_{s}=&&\frac{\hat{p}^{2}}{2M}-
\hat{\triangle}(\hat{\vec{p}},\vec{g})\hat{S}_{0}+\frac{1}{2}Mg^{2}t^{2}+\vec{g}.\hat{\vec{p}}t+
\lambda
(\sqrt{\hat{K}}\hat{S}_{-}+\sqrt{\hat{K}}\hat{S}_{+}) \\ \nonumber + && \omega_{c}\hat{K}-\frac{1}{2}\triangle
\hat{S}_{0}-\frac{q^{2}}{2M}\hat{S}_{0}+\frac{q^{2}}{8M}.
\end{eqnarray}
By using the same procedure as in [24], the Hamiltonian (13) takes the following form in the interaction picture
\begin{eqnarray}
\hat{\tilde{H}}^{I} _{s}=&&\omega_{c}(\hat{a}^{\dagger}\hat{a}+\frac{\hat{S}_{0}}{2})+\frac{1}{2}\hat{\Delta}(\hat{\vec{p}},\vec{g},t)\hat{S}_{0}\\ \nonumber +&& (\hat{\kappa}(t)
\sqrt{\hat{K}}\hat{S}_{-}+\hat{\kappa}^{*}(t)\sqrt{\hat{K}}\hat{S}_{+}),
\end{eqnarray}
where $\hat{\kappa}(t)$ is an effective coupling coefficient
\begin{equation}
\hat{\kappa}(t)= \lambda
\exp(\frac{it}{2}(\hat{\triangle}(\hat{\vec{p},t},\vec{g})+\frac{\hbar
q^{2} }{M})),
\end{equation}
 and the operator
\begin{equation}
\hat{\triangle}(\hat{\vec{p},t},\vec{g})=\omega_{c}-(\omega_{eg}+\frac{\vec{q}.\hat{\vec{p}}}{M}+\vec{q}.\vec{g}t+\frac{
q^{2}}{2M}),
\end{equation}
has been introduced as the Doppler shift detuning at time $t$
[23]. The Hamiltonian (14) has the form of the Hamiltonian of the
JCM, the only modification being the dependence of the detuning
on the conjugate momentum and the gravitational field.
Now according to ref.[23] we consider
\begin{equation}
\hat{\rho}_{sr}(t)=\exp(iM\vec{g}.\hat{\vec{x}}t)\hat{v}^{\dag} (\hat{\rho}_{sr})_{e} (t)  \hat{v},\hat{v}=\exp(-i\vec{q}.\hat{\vec{x}}\hat{S}_{0}).
\end{equation}
It can be shown that the operator $(\hat{\rho}_{sr})_{e}(t)$ satisfies an effective Schr\"{o}dinger
 equation governed by an effective
Hamiltonian $\hat{\tilde{H}}'_{s}=\hat{H}'_{s}+\hat{H}_{r}+\hat{V}_{sr}$, that is
\begin{equation}
i\frac{\partial(\hat{\rho}_{sr})_{e}(t)}{\partial
t}=\hat{\tilde{H}}'_{s}(\hat{\rho}_{sr})_{e}(t).
\end{equation}
By using the method given in [23]
we obtain the following equation of motion in the interaction picture
\begin{equation}
\frac{\partial\hat{\rho}_{sr}^{I} (t)}{\partial t}=-i[\hat{\tilde{H}}^{I} _{s},\hat{\rho}_{sr}^{I} (t)].
\end{equation}
The master equation for the reduced density operator of the atom-radiation system,
$\hat{\rho}_{s}^{I} (t)=Tr_{r}\hat{\rho}_{sr}^{I} (t)$, under the Markovian approximation
with neglecting the Lamb-shift term reads as [37]
\begin{equation}
\frac{\partial\hat{\rho}_{s}^{I} (t)}{\partial t}=-i\gamma[\hat{\tilde{H}}^{I} _{s},[\hat{\tilde{H}}^{I} _{s},\hat{\rho}_{s}^{I} (t_{0} )]].
\end{equation}
Now, we consider
\begin{equation}
\hat{\rho}_{s}(t)=\hat{U}_{1}^{\dag}\hat{\rho}_{s}^{I}\hat{U}_{1},
\end{equation}
with
\begin{equation}
\hat{U}_{1}=\textsf{T} \{\exp(i\int_{0}^{t}\hat{\tilde{H}}^{I} _{s}(t') dt')\},
\end{equation}
where the symbol $\textsf{T}$ denotes time ordering.
By using (14), (20), (21) and (22), and following the same procedure as in ref.[37] we obtain the master equation for the reduced density operator of the system under Markovian
approximation with neglecting the lamb shift term
\begin{equation}
\frac{\partial\hat{\rho}_{s}(t)}{\partial t}=-i[\hat{\tilde{H}}^{I} _{s},\hat{\rho}_{s}(t)]-\gamma [\hat{H}^{I} _{s},[\hat{H}^{I} _{s},\hat{\rho}_{s}(t)]],
\end{equation}
where $\hat{\tilde{H}}^{I} _{s}$ is given by (14). In Eq.(23), $\gamma$ is a parameter which depends on the temperature $T$
\begin{equation}
\gamma=\Delta\omega'+2\pi T lim_{\omega\rightarrow 0}(\frac{J(\omega)|\kappa(\omega)|^{2} }{\omega}),
\end{equation}
where
\begin{equation}
\Delta\omega'=i\int_{0}^{\infty}d\omega J(\omega) \frac{|\kappa(\omega)|^{2}}{\omega},
\end{equation}
and $J(\omega)$ and $\kappa(\omega)$ are the spectral density
 of the reservoir and the coupling coefficient, respectively. In the derivation of the master equation
 we have assumed that the parameter $T$ is high enough so that the Markovian approximation is valid.\\
\section{Dynamical Evolution of the Phase Damped JCM in the presence of Classical Gravity}
In section 2, we obtained the master equation
for the reduced density operator of the atom-radiation system under Markovian
approximation in the presence of a classical homogeneous gravitational
field. In this section,
We now start to find the exact solution for the density operator $\hat{\rho}_{s}(t)$ of the master equation (23)
with the Hamiltonian (14). For this purpose, we apply the approach presented in refs.[38-40].
The formal solution is given by
\begin{equation}
\hat{\rho}_{s}(t)=\exp(\hat{R}t)\exp(\hat{S}t)\exp(\hat{T}t)\hat{\rho}_{s}(0),
\end{equation}
where $\hat{\rho}_{s}(0)$ is the density operator of the initial atom-field system. The auxiliary
super-operators $\hat{R}$, $\hat{S}$ and $\hat{T}$ are defined through their action on the density operator such that
\begin{equation}
\exp(\hat{R}t)\hat{\rho}_{s}(0)\equiv \sum_{k=0}^{\infty}\frac{(2\gamma t)^{k}}{k!}(\hat{\tilde{H}}^{I}_{s})^{k}\hat{\rho}_{s}(0)(\hat{\tilde{H}}^{I}_{s})^{k},
\end{equation}
\begin{equation}
\exp(\hat{S}t)\hat{\rho}_{s}(0)\equiv \exp(-i\hat{\tilde{H}}^{I}_{s}t)\hat{\rho}_{s}(0)\exp(i\hat{\tilde{H}}^{I}_{s}t),
\end{equation}
\begin{equation}
\exp(\hat{T}t)\hat{\rho}_{s}(0)\equiv \exp(-\gamma (\hat{\tilde{H}}^{I}_{s})^{2} t)\hat{\rho}_{s}(0)\exp(-\gamma (\hat{\tilde{H}}^{I}_{s})^{2} t).
\end{equation}
\hspace*{00.5 cm}We assume that initially the radiation field is in a coherent
superposition of the Fock states, the atom is in the excited state $|e\rangle$, and the state
vector for the center-of-mass degree of freedom is
$|\psi_{c.m}(0)\rangle=\int d^{3}p \phi(\vec{p})|\vec{p}\rangle$.
Therefore, the initial density operator of the atom-radiation system reads as
\begin{equation}
\hat{\rho}_{s}(0)=\hat{\rho}_{field}(0)\otimes \hat{\rho}_{atom}(0)\otimes\hat{\rho}_{c.m}(0)= \left [\begin{array}{cccc}
\hat{\rho}_{field}(0)\otimes\hat{\rho}_{c.m}(0) & 0 \\
0 & 0 \\
\end{array}\right],
\end{equation}
where
\begin{equation}
\hat{\rho}_{field}(0)=\sum_{n}\sum_{m}w_{n}(0)w_{m}(0)|n\rangle \langle m |,
\end{equation}
\begin{equation}
\hat{\rho}_{c.m}(0)=\int d^{3}p \int d^{3}p'\phi^{*}(\vec{p'})\phi(\vec{p})|\vec{p}\rangle \langle\vec{p'} |,
\end{equation}
with $w_{n}(0)=\frac{\exp(-\frac{|\alpha|^{2}}{2})\alpha^{n}}{\sqrt{n!}}$.
The Hamiltonian (14) can be expressed as a sum of two terms which commute with each other, that is,
\begin{equation}
\hat{\tilde{H}}^{I}_{s}=\hat{H}_{1}+\hat{H}_{2},[\hat{H}_{1},\hat{H}_{2}]=0
\end{equation}
where
\begin{equation}
\hat{H}_{1}=\omega_{c}(\hat{a}^{\dagger}\hat{a}+\frac{\hat{S}_{0}}{2}),
\end{equation}
\begin{equation}
\hat{H}_{2}=\frac{1}{2}\hat{\Delta}(\hat{\vec{p}},\vec{g},t)\hat{S}_{0}+ (\hat{\kappa}(t)
\sqrt{\hat{K}}\hat{S}_{-}+\hat{\kappa}^{*}(t)\sqrt{\hat{K}}\hat{S}_{+}).
\end{equation}
In the two-dimensional atomic basis we have
\begin{equation}
\hat{H}_{1}=\omega_{c} \left [\begin{array}{cccc}
\hat{n}+\frac{1}{2} & 0 \\
0 & \hat{n}-\frac{1}{2} \\
\end{array}\right],
\end{equation}
\begin{equation}
\hat{H}_{2}= \left [\begin{array}{cccc}
\frac{\Delta(\vec{p},\vec{g},t)}{4} & \kappa^{*}(t)\hat{a}  \\
\kappa(t) \hat{a}^{\dagger} & -\frac{\Delta(\vec{p},\vec{g},t)}{4} \\
\end{array}\right].
\end{equation}
Also, the square of the Hamiltonian (14) can be expressed as a sum of two operators, one of them is diagonal, in the form
\begin{equation}
(\hat{\tilde{H}}^{I}_{s})^{2}=\hat{A}_{1}+\hat{A}_{2},
\end{equation}
where
\begin{eqnarray}
\hat{A}_{1}=&&\hat{H}_{1}^{2}+\hat{H}_{2}^{2} \\ \nonumber  =&&\left [\begin{array}{cccc}
\omega_{c}^{2}(\hat{n}+\frac{1}{2})^{2}+\lambda^{2}(\hat{n}+1)+(\frac{\Delta(\vec{p},\vec{g},t)}{4})^{2} & 0 \\
0 & \omega_{c}^{2}(\hat{n}-\frac{1}{2})^{2}+\lambda^{2}\hat{n}+(\frac{\Delta(\vec{p},\vec{g},t)}{4})^{2} \\
\end{array}\right],
\end{eqnarray}
and
\begin{equation}
\hat{A}_{2}=2\hat{H}_{1}\hat{H}_{2}=2\omega_{c} \left [\begin{array}{cccc}
(\hat{n}+\frac{1}{2})(\frac{\Delta(\vec{p},\vec{g},t)}{4}) & (\hat{n}+\frac{1}{2})\kappa^{*}(t)\hat{a} \\
(\hat{n}-\frac{1}{2})\kappa(t) \hat{a}^{\dagger} & -(\hat{n}-\frac{1}{2})(\frac{\Delta(\vec{p},\vec{g},t)}{4}) \\
\end{array}\right].
\end{equation}
It is easily proved that $[\hat{A}_{1},\hat{A}_{2}]=0$. Taking into account the initial condition (30)
we define the auxiliary density operator $\hat{\rho}_{2}(t)$ as
\begin{eqnarray}
\hat{\rho}_{2}(t)&&=\exp(\hat{S}t)\exp(\hat{T}t)\hat{\rho}_{s}(0)\\ \nonumber &&=\exp(-i\hat{H}_{2}t)\exp(-\gamma \hat{A}_{2} t)\hat{\rho}_{1}(t)\exp(-\gamma \hat{A}_{2} t)\exp(i\hat{H}_{2}t),
\end{eqnarray}
where the operator $\hat{\rho}_{1}(t)$ is defined by
\begin{equation}
\hat{\rho}_{1}(t)=|\Psi(t)\rangle\langle\Psi(t)|\otimes |e\rangle\langle e|,
\end{equation}
with
\begin{equation}
|\Psi(t)\rangle=\exp(-\gamma t[\omega_{c}^{2}(\hat{n}+\frac{1}{2})^{2}+\lambda^{2}(\hat{n}+1)+(\frac{\Delta(\vec{p},\vec{g},t)}{4})^{2}]) w_{n}(0)\exp(-in\omega_{c}t)|n\rangle.
\end{equation}
From (36) and (39) we have, respectively
\begin{equation}
\exp(-i \hat{H}_{1}t)= \left [\begin{array}{cccc}
 \exp(-i\omega_{c}(\hat{n}+\frac{1}{2}))& 0 \\
0 & \exp(-i\omega_{c}(\hat{n}-\frac{1}{2})) \\
\end{array}\right],
\end{equation}
\begin{equation}
\exp(-\gamma \hat{A}_{1}t)= \left [\begin{array}{cccc}
 (\hat{A}_{1})_{11}(\hat{n},t)& 0 \\
0 & (\hat{A}_{1})_{22}(\hat{n},t) \\
\end{array}\right],
\end{equation}
where
\begin{equation}
(\hat{A}_{1})_{11}(\hat{n},t)=\exp(-\gamma t[\omega_{c}^{2}(\hat{n}+\frac{1}{2})^{2}+\lambda^{2}(\hat{n}+1)+(\frac{\Delta(\vec{p},\vec{g},t)}{4})^{2}]),
\end{equation}
\begin{equation}
(\hat{A}_{1})_{22}(\hat{n},t)=\exp(-\gamma t[\omega_{c}^{2}(\hat{n}-\frac{1}{2})^{2}+\lambda^{2}\hat{n}+(\frac{\Delta(\vec{p},\vec{g},t)}{4})^{2}]).
\end{equation}
Also, we can write
\begin{equation}
\exp(-\gamma \hat{A}_{2}t)= \left [\begin{array}{cccc}
\hat{e}_{1}(\hat{n},t)& \hat{e}_{2}(\hat{n},t)\hat{a} \\
 \hat{e}_{3}(\hat{n},t)\hat{a}^{\dagger}& \hat{e}_{4}(\hat{n},t) \\
\end{array}\right],
\end{equation}
where
\begin{equation}
\hat{e}_{1}(\hat{n},t)=\cos(\gamma t \sqrt{\hat{c}_{1}(\hat{n},t)} )-\omega_{c}(\frac{\Delta(\vec{p},\vec{g},t)}{2})(\hat{n}+\frac{1}{2} )\frac{\sinh(\gamma t \sqrt{\hat{c}_{1}(\hat{n},t)})}{\sqrt{\hat{c}_{1}(\hat{n},t)}},
\end{equation}
\begin{equation}
\hat{e}_{2}(\hat{n},t)=-2\omega_{c}\lambda(\hat{n}-\frac{1}{2} )\frac{\sinh(\gamma t \sqrt{\hat{c}_{1}(\hat{n}-1,t)})}{\sqrt{\hat{c}_{1}(\hat{n}-1,t)}},
\end{equation}
\begin{equation}
\hat{e}_{3}(\hat{n},t)=-2\omega_{c}\lambda(\hat{n}-\frac{1}{2} )\frac{\sinh(\gamma t \sqrt{\hat{c}_{2}(\hat{n},t)})}{\sqrt{\hat{c}_{2}(\hat{n},t)}},
\end{equation}
\begin{equation}
\hat{e}_{4}(\hat{n},t)=\cos(\gamma t \sqrt{\hat{c}_{2}(\hat{n},t)} )-\omega_{c}(\frac{\Delta(\vec{p},\vec{g},t)}{2})(\hat{n}-\frac{1}{2} )\frac{\sinh(\gamma t \sqrt{\hat{c}_{2}(\hat{n},t)})}{\sqrt{\hat{c}_{2}(\hat{n},t)}},
\end{equation}
with
\begin{equation}
\hat{c}_{1}(\hat{n},t)=\omega_{c}^{2}(\frac{\Delta(\vec{p},\vec{g},t)}{2})^{2}(\hat{n}+\frac{1}{2})^{2}+\lambda^{2}(\frac{\Delta(\vec{p},\vec{g},t)}{2})^{2}(\hat{n}+1)(\hat{n}+\frac{1}{2})^{2},
\end{equation}
\begin{equation}
\hat{c}_{2}(\hat{n},t)=\omega_{c}^{2}(\frac{\Delta(\vec{p},\vec{g},t)}{2})^{2}(\hat{n}-\frac{1}{2})^{2}+\lambda^{2}(\frac{\Delta(\vec{p},\vec{g},t)}{2})^{2}\hat{n}(\hat{n}-\frac{1}{2})^{2}.
\end{equation}
Similarly, we can express the operator $\exp(-i \hat{H}_{2}t)$ in the two-dimensional atomic basis as
\begin{equation}
\exp(-i \hat{H}_{2}t)= \left [\begin{array}{cccc}
\hat{d}_{1}(\hat{n},t)&\hat{d}_{2}(\hat{n},t)\hat{a} \\
 \hat{d}_{3}(\hat{n},t)\hat{a}^{\dagger}& \hat{d}_{4}(\hat{n},t) \\
\end{array}\right],
\end{equation}
where
\begin{equation}
\hat{d}_{1}(\hat{n},t)=\cos(t((\frac{\Delta(\vec{p},\vec{g},t)}{4})^{2} +\lambda^{2}(\hat{n}+1)))-(\frac{\Delta(\vec{p},\vec{g},t)}{4})\frac{\sin(t((\frac{\Delta(\vec{p},\vec{g},t)}{4})^{2} +\lambda^{2}(\hat{n}+1)))}{\sqrt{(\frac{\Delta(\vec{p},\vec{g},t)}{4})^{2} +\lambda^{2}(\hat{n}+1))}},
\end{equation}
\begin{equation}
\hat{d}_{2}(\hat{n},t)=-i\lambda \frac{\sin(t((\frac{\Delta(\vec{p},\vec{g},t)}{4})^{2} +\lambda^{2}(\hat{n}+1)))}{\sqrt{(\frac{\Delta(\vec{p},\vec{g},t)}{4})^{2} +\lambda^{2}(\hat{n}+1)}},
\end{equation}
\begin{equation}
\hat{d}_{3}(\hat{n},t)=-i\lambda \frac{\sin(t((\frac{\Delta(\vec{p},\vec{g},t)}{4})^{2} +\lambda^{2}\hat{n}))}{\sqrt{(\frac{\Delta(\vec{p},\vec{g},t)}{4})^{2} +\lambda^{2}\hat{n}}},
\end{equation}
\begin{equation}
\hat{d}_{4}(\hat{n},t)=\cos(t((\frac{\Delta(\vec{p},\vec{g},t)}{4})^{2} +\lambda^{2}\hat{n}))-(\frac{\Delta(\vec{p},\vec{g},t)}{4})\frac{\sin(t((\frac{\Delta(\vec{p},\vec{g},t)}{4})^{2} +\lambda^{2}\hat{n}))}{\sqrt{(\frac{\Delta(\vec{p},\vec{g},t)}{4})^{2} +\lambda^{2}\hat{n})}}.
\end{equation}
Then, from (48) and (55), it follows that
\begin{equation}
\exp(-i \hat{H}_{2}t)\exp(-\gamma \hat{A}_{2}t)= \left [\begin{array}{cccc}
\hat{f}_{1}(\hat{n},t)&\hat{f}_{2}(\hat{n},t)\hat{a} \\
 \hat{f}_{3}(\hat{n},t)\hat{a}^{\dagger}& \hat{f}_{4}(\hat{n},t) \\
\end{array}\right],
\end{equation}
where
\begin{equation}
\hat{f}_{1}(\hat{n},t)=\hat{e}_{1}(\hat{n},t)\hat{d}_{1}(\hat{n},t)+\hat{e}_{2}(\hat{n},t)\hat{d}_{2}(\hat{n},t),
\end{equation}
\begin{equation}
\hat{f}_{2}(\hat{n},t)=\hat{e}_{2}(\hat{n},t)\hat{d}_{1}(\hat{n},t)+\hat{e}_{1}(\hat{n},t)\hat{d}_{2}(\hat{n},t),
\end{equation}
\begin{equation}
\hat{f}_{3}(\hat{n},t)=\hat{e}_{3}(\hat{n},t)\hat{d}_{4}(\hat{n},t)+\hat{e}_{4}(\hat{n},t)\hat{d}_{3}(\hat{n},t),
\end{equation}
\begin{equation}
\hat{f}_{4}(\hat{n},t)=\hat{e}_{4}(\hat{n},t)\hat{d}_{4}(\hat{n},t)+\hat{e}_{3}(\hat{n},t)\hat{d}_{3}(\hat{n},t).
\end{equation}
Substituting (42) and (60) into (41), we can obtain an explicit expression for the operator $\hat{\rho}_{2}(t)$ as follows
\begin{equation}
(\hat{\rho}_{2}(t))_{i,j}=|\Psi_{i}(t)\rangle\langle\Psi_{j}(t)|,(i,j=1,2),
\end{equation}
with
\begin{equation}
|\Psi_{1}(t)\rangle=\hat{f}_{1}(\hat{n},t)|\Psi(t)\rangle, |\Psi_{2}(t)\rangle=\hat{f}_{3}(\hat{n},t)|\Psi(t)\rangle,
\end{equation}
where $|\Psi(t)\rangle$ is given by Eq.(43).
Now, we obtain the action of the operator $\exp(\hat{R}t)$ on the operator $\hat{\rho}_{2}(t)$
\begin{equation}
\hat{\rho}_{s}(t)=\sum_{k=0}^{\infty}\frac{(2 \gamma t)^{k} }{k!}\hat{H}^{k}\hat{\rho}_{2}(t)\hat{H}^{k},
\end{equation}
where
\begin{equation}
\hat{H}^{k}=\sum_{l=0}^{k}\frac{k! }{l!(k-l)!}\hat{H}_{1}^{k-l}\hat{H}_{2}^{l},
\end{equation}
which can be explicitly expressed as follows
\begin{equation}
\hat{H}^{k}= \left [\begin{array}{cccc}
\hat{g}_{+}^{k} (\hat{n},t)&\kappa^{*}(t) \frac{\hat{u}_{-}^{k} (\hat{n},t)}{\sqrt{(\frac{\Delta(\vec{p},\vec{g},t)}{4})^{2} +\lambda^{2}(\hat{n}+1))}} \hat{a} \\
\kappa(t)\frac{\hat{v}_{-}^{k} (\hat{n},t)}{\sqrt{(\frac{\Delta(\vec{p},\vec{g},t)}{4})^{2} +\lambda^{2}(\hat{n}+1))}} \hat{a}^{\dagger}&\hat{g}_{-}^{k} (\hat{n},t)  \\
\end{array}\right],
\end{equation}
where
\begin{equation}
\hat{g}_{+}^{k} (\hat{n},t)=\hat{u}_{+}^{k} (\hat{n},t)+\frac{\Delta(\vec{p},\vec{g},t)}{4}\hat{u}_{-}^{k} (\hat{n},t),
\end{equation}
\begin{equation}
\hat{g}_{-}^{k} (\hat{n},t)=\hat{v}_{+}^{k} (\hat{n},t)-\frac{\Delta(\vec{p},\vec{g},t)}{4}\hat{v}_{-}^{k} (\hat{n},t),
\end{equation}
\begin{equation}
\hat{u}_{\pm}^{k} (\hat{n},t)=\frac{1}{2}(\hat{r}_{+}^{k} (\hat{n},t) \pm \hat{r}_{-}^{k} (\hat{n},t)),\hat{v}_{\pm}^{k} (\hat{n},t)=\frac{1}{2}(\hat{s}_{+}^{k} (\hat{n},t) \pm \hat{s}_{-}^{k} (\hat{n},t)),
\end{equation}
with
\begin{equation}
\hat{r}_{\pm} (\hat{n},t)=\omega_{c}(\hat{n}+\frac{1}{2} )\pm\sqrt{(\frac{\Delta(\vec{p},\vec{g},t)}{4})^{2} +\lambda^{2}(\hat{n}+1))},
\end{equation}
\begin{equation}
\hat{s}_{\pm} (\hat{n},t)=\omega_{c}(\hat{n}-\frac{1}{2} )\pm\sqrt{(\frac{\Delta(\vec{p},\vec{g},t)}{4})^{2} +\lambda^{2}\hat{n}}.
\end{equation}
Finally, by substituting (65) and (69) into (67) we obtain the exact solution of the master equation (23) for the phase
damped JCM in the presence of a classical homogeneous gravity field
\begin{equation}
\hat{\rho}_{s}(t)= \left [\begin{array}{cccc}
 \sum_{k=0}^{\infty}\frac{(2 \gamma t)^{k} }{k!}\hat{M}_{11} ^{k}(t) &\sum_{k=0}^{\infty}\frac{(2 \gamma t)^{k} }{k!}\hat{M}_{12} ^{k}(t) \\
 \sum_{k=0}^{\infty}\frac{(2 \gamma t)^{k} }{k!}\hat{M}_{21} ^{k}(t) &  \sum_{k=0}^{\infty}\frac{(2 \gamma t)^{k} }{k!}\hat{M}_{22} ^{k}(t)\\
\end{array}\right],
\end{equation}
where
\begin{eqnarray}
\hat{M}_{11} ^{k}(t) =&&(\hat{g}_{+}^{k} (\hat{n},t)\hat{\Psi}_{11}(t)\hat{g}_{+}^{k} (\hat{n},t)+\hat{a}\hat{v}_{-}^{'k} (\hat{n},t)\hat{\Psi}_{21}(t)\hat{g}_{+}^{k} (\hat{n},t)\\ \nonumber
+&&\hat{g}_{+}^{k} (\hat{n},t)\hat{\Psi}_{12}(t)\hat{v}_{-}^{'k} (\hat{n},t)\hat{a}^{\dagger}+\hat{a}\hat{v}_{-}^{'k} (\hat{n},t)\hat{\Psi}_{22}(t)\hat{v}_{-}^{'k} (\hat{n},t)\hat{a}^{\dagger})|\phi(\vec{p})|^{2},
\end{eqnarray}
\begin{eqnarray}
\hat{M}_{22} ^{k}(t) =&&(\hat{v}_{-}^{'k} (\hat{n},t)\hat{a}^{\dagger}\hat{\Psi}_{11}(t)\hat{a}\hat{v}_{-}^{'k} (\hat{n},t)+\hat{g}_{-}^{k} (\hat{n},t)\hat{\Psi}_{21}(t)\hat{a}\hat{v}_{+}^{'k} (\hat{n},t)\\ \nonumber
+&&\hat{v}_{-}^{'k} (\hat{n},t)\hat{a}^{\dagger}\hat{\Psi}_{12}(t)\hat{g}_{-}^{k} (\hat{n},t)+\hat{g}_{-}^{k} (\hat{n},t)\hat{\Psi}_{22}(t)\hat{g}_{-}^{k} (\hat{n},t))|\phi(\vec{p})|^{2},
\end{eqnarray}
\begin{eqnarray}
\hat{M}_{21}^{k}(t)&&=(\hat{M}_{12}^{k}(t))^{\dagger}=(\hat{v}_{-}^{'k} (\hat{n},t)\hat{a}^{\dagger}\hat{\Psi}_{11}(t)\hat{g}_{+}^{k} (\hat{n},t)+\hat{g}_{-}^{k} (\hat{n},t)\hat{\Psi}_{21}(t)\\ \nonumber && \times \hat{a}\hat{g}_{+}^{k} (\hat{n},t)
+\hat{v}_{-}^{'k} (\hat{n},t)\hat{a}^{\dagger}\hat{\Psi}_{12}(t)\hat{v}_{-}^{'k} (\hat{n},t)\hat{a}^{\dagger}+\hat{g}_{-}^{k} (\hat{n},t)\hat{\Psi}_{22}(t)\hat{v}_{-}^{'k} (\hat{n},t)\hat{a}^{\dagger})|\phi(\vec{p})|^{2},
\end{eqnarray}
with
\begin{equation}
\hat{v}_{-}^{'k} (\hat{n},t)=\frac{\lambda}{\sqrt{(\frac{\Delta(\vec{p},\vec{g},t)}{4} )^{2}+ \lambda^{2}\hat{n} }}\hat{v}_{-}^{k} (\hat{n},t).
\end{equation}
\hspace*{00.5 cm}Making use of the solution given by (75), one can evaluate the mean values of operators of interest. In the next section
we shall use it to investigate various dynamical properties of the phase damped JCM in the presence of a homogeneous
gravitational field.
\section{Dynamical Properties}
In this section, we study the influence of the gravitational
field on the quantum statistical properties of the atom and the
 quantized radiation field in the presence of the phase damping.  \\
 \\
 \\
 \\
 \\
{\bf  4a. Atomic Population Inversion} \\
\\
An important quantity is the  atomic population inversion which
is expressed by the expression
\begin{equation}
W(t)=\langle\hat{\sigma}_{3}(t)\rangle=Tr_{atom}(\hat{\rho}_{atom}(t)\hat{\sigma}_{3}(t)),
\end{equation}
where
\begin{equation}
\hat{\rho}_{atom}(t)=Tr_{field}(\hat{\rho}_{s}(t).
\end{equation}
We can rewrite (80) as follows
\begin{equation}
W(t)=\int d^{3}p\sum_{i=e,g}\langle i|\hat{\rho}_{atom}(t)\hat{\sigma}_{3}(t)|i \rangle=\int d^{3}p\sum_{n=0}^{\infty}(\langle n|\otimes(\langle e|\hat{\rho}_{s}(t)|e \rangle- \langle g|\hat{\rho}_{s}(t)|g \rangle    )\otimes|n \rangle ).
\end{equation}
Therefore, by using (75) and (82) we obtain
\begin{equation}
W(t)=\int d^{3}p (\sum_{k=0}^{\infty}\sum_{n=0}^{\infty} \frac{(2 \gamma t)^{k} }{k!}(\langle n | \hat{M}_{11} ^{k}(t)|n \rangle -\langle n | \hat{M}_{22} ^{k}(t)|n \rangle )),
\end{equation}
where from (76) and (77) we have
\begin{eqnarray}
\langle n | \hat{M}_{11} ^{k}(t)|n \rangle &&= (g_{+}^{k} (n,t))^{2}|\psi_{1}(n,t) |^{2}+(h^{k} (n+1,t))^{2}\\
\nonumber && \times|\psi_{2}(n+1,t) |^{2}   +  2Re[g_{+}^{k} (n,t)h^{k} (n+1,t)\psi_{1}^{*} (n,t)\psi_{2}(n+1,t) ],
\end{eqnarray}
\begin{eqnarray}
\langle n | \hat{M}_{22} ^{k}(t)|n \rangle &&= (h^{k} (n,t))^{2}|\psi_{1}(n-1,t) |^{2}+(g_{-} ^{k} (n,t))^{2}|\psi_{2}(n,t) |^{2}  \\
\nonumber &&+  2Re[g_{-}^{k} (n,t)h^{k} (n,t)\psi_{1}^{*} (n-1,t)\psi_{2}(n,t) ],
\end{eqnarray}
with
\begin{equation}
h^{k} (n,t)=\sqrt{n}v_{-}^{' k} (n,t),
\end{equation}
and
\begin{equation}
 \psi_{i}(n,t)=\langle n|\Psi_{i}(t)\rangle,(i=1,2),
\end{equation}
where we have defined $|\Psi_{i}(t)\rangle$ in (66).\\
 \hspace*{00.5 cm} In Fig.1 we have plotted the atomic population inversion as a
 function of the scaled time $\lambda t$ for three different values of the
 parameter $\vec{q}.\vec{g}$. In this figure and all the
 subsequent figures we set
$q=10^{7}m^{-1}$, $M=10^{-26}Kg$, $g=9.8\frac{m}{s^{2}}$,
$\omega_{rec}=\frac{\hbar q^{2}}{2M}=.5\times10^{6}\frac{rad}{s}$,
$\lambda=1\times 10^{6}\frac{rad}{s}$,
$\triangle_{0}=8.5\times10^{7}\frac{rad}{s}$, $ \alpha=2$,
$\Delta=1.8\times 10^{6}\frac{rad}{s}$,
$\phi(\vec{p})=\frac{1}{\sqrt{2\pi
\sigma_{0}}}\exp(\frac{-p^{2}}{\sigma_{0}^{2}})$ with
$\sigma_{0}=1$ [21-23] and $\gamma=7 \times 10^{-5}\frac{rad}{s} $. It should be noted that the relevant
time scale introduced by the gravitational influence is $\tau_{a}=\frac{1}{\sqrt{\vec{q}.\vec{g} } } $ [23].
For an optical laser with $q=10^{7}m^{-1}$, $\tau_{a}$
is about $10^{-4}s$.
In Fig.1a we consider small gravitational
influence in the presence of the phase damping. This means very small $\vec{q}.\vec{g}$, i.e., the
momentum transfer from the laser beam to the atom is only
slightly altered by the gravitational acceleration because the
latter is very small or nearly perpendicular to the laser beam.
In Figs.1b and 1c we consider the gravitational influence in the presence of the phase damping for
$\vec{q}.\vec{g}=0.5\times 10^{7}$ and $\vec{q}.\vec{g}=1.5\times
10^{7}$, respectively. By comparing Figs.1a, 1b and 1c we can see
the influence of gravity on the time evolution of the atomic
population inversion when there is the phase damping. As it is seen from Fig.1a for the atomic
population inversion the Rabi-like oscillations can be
identified. With the increasing value of the parameter
$\vec{q}.\vec{g}$ (see Figs.1b and 1c) the Rabi oscillations of
the atomic population inversion disappear.\\
\\
{\bf  4b. Atomic Dipole Squeezing} \\
\\
To analyze the quantum fluctuations of the atomic dipole variables and
examine their squeezing we consider the two slowly varying
Hermitian quadrature operators
\begin{equation}
\hat{\sigma}_{1}=\frac{1}{2}(\hat{\sigma}_{+}\exp(-i\omega_{eg}t)+\hat{\sigma}_{-}\exp(i\omega_{eg}t)),
\end{equation}
and
\begin{equation}
\hat{\sigma}_{2}=\frac{1}{2i}(\hat{\sigma}_{+}\exp(-i\omega_{eg}t)-\hat{\sigma}_{-}\exp(i\omega_{eg}t)).
\end{equation}
In fact $\hat{\sigma}_{1}$ and $\hat{\sigma}_{2}$ correspond to
the dispersive and absorptive components of the amplitude of the
atomic polarization [41], respectively. They obey the commutation
relation
$[\hat{\sigma}_{1},\hat{\sigma}_{2}]=\frac{i}{2}\hat{\sigma}_{3}$.
Correspondingly, the Heisenberg uncertainty relation is
\begin{equation}
(\Delta \hat{\sigma}_{1} )^{2} (\Delta \hat{\sigma}_{2} )^{2}\geq
\frac{1}{16}|\langle\hat{\sigma}_{3} \rangle|^{2},
\end{equation}
where $(\Delta \hat{\sigma}_{i} )^{2}=\langle \hat{\sigma}_{i}
^{2} \rangle - \langle \hat{\sigma}_{i} \rangle ^{2}$ is the
variance in the component $\hat{\sigma}_{i}(i=1,2)$ of the atomic
dipole. \\
 \hspace*{00.5 cm}The fluctuations in the component
 $\hat{\sigma}_{i}(i=1,2)$are said to be squeezed (i.e., dipole
 squeezing)if the variance in $\hat{\sigma}_{i}$ satisfies the
 condition
\begin{equation}
(\Delta \hat{\sigma}_{i} )^{2}<\frac{1}{4}|\langle
\hat{\sigma}_{3}\rangle|, (i=1 or 2).
\end{equation}
Since $\hat{\sigma}_{i}^{2}=\frac{1}{4}$ this condition may be
written as
\begin{equation}
F_{i}=1-4 \langle \hat{\sigma}_{i} \rangle ^{2}-|\langle
\hat{\sigma}_{3}\rangle|<0, (i=1 or 2).
\end{equation}
The expectation values of the atomic operators $\hat{\sigma}_{+}$
and $\hat{\sigma}_{-}$ are given by
\begin{equation}
\langle\hat{\sigma}_{\pm}(t)\rangle=Tr_{atom}(\hat{\rho}_{atom}(t)\hat{\sigma}_{\pm}(t)).
\end{equation}
We can rewrite (93) as follows
\begin{equation}
\langle\hat{\sigma}_{\pm}(t)\rangle=\int d^{3}p\sum_{n=0}^{\infty}(\langle n|\otimes(\langle e|\hat{\rho}_{s}(t)\hat{\sigma}_{\pm}(t)|e \rangle- \langle g|\hat{\rho}_{s}(t)\hat{\sigma}_{\pm}(t)|g \rangle    )\otimes|n \rangle ).
\end{equation}
Therefore, by using (75) and (94) we obtain
\begin{equation}
\langle\hat{\sigma}_{-}(t)\rangle= \int d^{3}p\sum_{k=0}^{\infty}\sum_{n=0}^{\infty} \frac{(2 \gamma t)^{k} }{k!}\langle n | \hat{M}_{12} ^{k}(t)|n \rangle =\langle\hat{\sigma}_{+}(t)\rangle^{*},
\end{equation}
where from (78) we have
\begin{eqnarray}
\langle n | \hat{M}_{12} ^{k}(t)|n \rangle &&= g_{+}^{k} (n,t)\psi_{1}(n,t) ( g_{-}^{k} (n,t)\psi_{2}^{*} (n,t)+h^{k} (n,t)\psi_{1}^{*} (n-1,t))\\
\nonumber && +h^{k} (n+1,t)\psi_{2}(n+1,t)(h^{k} (n,t)\psi_{1}^{*} (n-1,t)+g_{-}^{k} (n,t)\psi_{2}^{*} (n,t)).
\end{eqnarray}
 \hspace*{00.5 cm}The time evolution of $F_{1}(t)$ corresponding to the squeezing
of $\hat{\sigma}_{1}$ has been shown in Fig.2 for three values of
the parameter $\vec{q}.\vec{g}$ in the presence of the phase damping. As it is seen, with the
increasing value of the parameter $\vec{q}.\vec{g}$ the dipole
squeezing is completely removed.\\
\\
\\
{\bf  4c. Atomic momentum diffusion} \\
 \\
 The next quantity we examine is the atomic momentum diffusion. As a consequence of the atomic momentum
 diffusion, the atom experiences light-induced forces (radiation force) during its
 interaction with the radiation field.
The atomic momentum diffusion is given
 by
\begin{equation}
\Delta p(t)=(\langle \hat{p}(t)^{2}\rangle - \langle \hat{p}(t)
\rangle^{2} )^{\frac{1}{2}}.
\end{equation}
Now we calculate the expectation values of the $\hat{\vec{p}}$ and $\hat{\vec{p}}^{2} $
\begin{equation}
\langle\hat{\vec{p}}(t)\rangle=Tr_{atom}(\hat{\rho}_{atom}(t)\hat{\vec{p}}(t)),
\end{equation}
\begin{equation}
\langle\hat{\vec{p}}(t)^{2} \rangle=Tr_{atom}(\hat{\rho}_{atom}(t)\hat{\vec{p}}(t)^{2} ).
\end{equation}
By using $\hat{p}|p\rangle=p|p\rangle$ and (75) we obtain
\begin{equation}
\langle\hat{\vec{p}}(t)\rangle=\int d^{3}p\sum_{k=0}^{\infty}\sum_{n=0}^{\infty} \frac{(2 \gamma t)^{k} }{k!}p(\langle n | \hat{M}_{11} ^{k}(t)|n \rangle -\langle n | \hat{M}_{22} ^{k}(t)|n \rangle ),
\end{equation}
\begin{equation}
\langle\hat{\vec{p}}(t)^{2} \rangle=\int d^{3}p\sum_{k=0}^{\infty}\sum_{n=0}^{\infty} \frac{(2 \gamma t)^{k} }{k!}p^{2} (\langle n | \hat{M}_{11} ^{k}(t)|n \rangle -\langle n | \hat{M}_{22} ^{k}(t)|n \rangle ),
\end{equation}
where $\langle n | \hat{M}_{11} ^{k}(t)|n \rangle$ and $\langle n | \hat{M}_{22} ^{k}(t)|n \rangle$ are given by (84) and (85), respectively.\\
\hspace*{00.5 cm}In Figs. 3a-3b we have plotted $\Delta p(t)$ for $\vec{q}.\vec{g}=0$,
$\vec{q}.\vec{g}=0.5\times10^{7}$ and
$\vec{q}.\vec{g}=1.5\times10^{7}$, respectively. In figure 3a the
Rabi-like oscillations can be identified, but in Figs.3b and 3c,
when the influence of the gravitational field increases, the Rabi
oscillations disappear and the atomic momentum diffusion becomes positive and the atom does not recoil, because
before absorption of photon, the atom is deflected by the gravitational field . Moreover, the atom can experience larger
light-induced forces during
 its
 interaction with the radiation field, when the gravitational field
 increases.\\
  \\
{\bf  4d. Photon Counting Statistics} \\
\\
 We now investigate the influence of gravity on the sub-Poissonian statistics of the radiation
 field. For this purpose, we calculate the Mandel parameter defined by [42]
\begin{equation}
Q(t)=\frac{(\langle n(t)^{2}\rangle-\langle n(t)\rangle
^{2})}{\langle n(t)\rangle}-1.
\end{equation}
For $Q<0$ $(Q>0)$, the statistics is sub-Poissonian
(super-Poissonian); $Q=0$ stands for Poissonian statistics. Since
$\langle n(t)\rangle=\sum_{n=0}^{\infty}n P(n,t)$ and $\langle
n(t)^{2}\rangle=\sum_{n=0}^{\infty}n^{2} P(n,t)$ we have
\begin{equation}
Q(t)=(\{[\sum_{n=0}^{\infty} n^{2}P(n,t)
]-[\sum_{n=0}^{\infty} n P(n,t) ]^{2}    \}
[\sum_{n=0}^{\infty} n P(n,t) ]^{-1})-1,
\end{equation}
where the probability of finding $n$ photons in the radiation field is found to be
\begin{equation}
P(n,t)=\langle n|\hat{\rho}_{field}(t)|n \rangle=\langle n|Tr_{atom} \hat{\rho}_{s}(t)|n \rangle,
\end{equation}
and by using (75) we have
\begin{equation}
P(n,t)=\int d^{3}p\sum_{k=0}^{\infty}\sum_{n=0}^{\infty} \frac{(2 \gamma t)^{k} }{k!}(\langle n | \hat{M}_{11} ^{k}(t)|n \rangle +\langle n | \hat{M}_{22} ^{k}(t)|n \rangle ).
\end{equation}
Therefore, by using (103) and (105) we obtain
\begin{eqnarray}
Q(t)&&=(\{[\sum_{n=0}^{\infty} n^{2}(\int d^{3}p\sum_{k=0}^{\infty}\sum_{n=0}^{\infty} \frac{(2 \gamma t)^{k} }{k!}(\langle n | \hat{M}_{11} ^{k}(t)|n \rangle +\langle n | \hat{M}_{22} ^{k}(t)|n \rangle ))
]\\ \nonumber &&-[\sum_{n=0}^{\infty} n (\int d^{3}p\sum_{k=0}^{\infty}\sum_{n=0}^{\infty} \frac{(2 \gamma t)^{k} }{k!}(\langle n | \hat{M}_{11} ^{k}(t)|n \rangle +\langle n | \hat{M}_{22} ^{k}(t)|n \rangle )) ]^{2}    \}
\\ \nonumber &&\times[\sum_{n=0}^{\infty} n (\int d^{3}p\sum_{k=0}^{\infty}\sum_{n=0}^{\infty} \frac{(2 \gamma t)^{k} }{k!}(\langle n | \hat{M}_{11} ^{k}(t)|n \rangle +\langle n | \hat{M}_{22} ^{k}(t)|n \rangle )) ]^{-1})-1.
\end{eqnarray}
\hspace*{00.5 cm}The numerical results for three values of the parameter
$\vec{q}.\vec{g}$ are shown in Fig.4. As
it is seen, the cavity-field exhibits alternately sub-Poissonian and
super-Poissonian statistics when the influence of the
gravitational field is negligible. With increasing
$\vec{q}.\vec{g}$ the sub-Poissonian characteristic is suppressed
and the cavity-field exhibits super-Poissonian statistics.
 After some time, the Mandel parameter $Q$ is stabilized at an asymptotic
zero value; the larger the parameter $\vec{q}.\vec{g}$ is more rapidly $Q(t)$ reaches the asymptotic value zero.\\
\\
{\bf  4e. Quadrature Squeezing of the Cavity-Field} \\
\\
 Finally, we investigate the influence of gravity on the quadrature squeezing of the radiation field.
  For this purpose, we introduce
 two slowly varying quadrature operators
\begin{equation}
\hat{X}_{1}(t)=\frac{1}{2}(\hat{a}\exp(i \omega t
)+\hat{a}^{\dagger}\exp(-i\omega t )),
\end{equation}
and
\begin{equation}
\hat{X}_{2}(t)=\frac{1}{2i}(\hat{a}\exp(i\omega t)
-\hat{a}^{\dagger}\exp(-i\omega t)),
\end{equation}
where $\hat{a}$ and $\hat{a}^{\dagger}$ obey the commutation
relation $[ \hat{a},\hat{a}^{\dagger}  ]=1$. The operators
$\hat{X}_{1}(t)$ and $\hat{X}_{2}(t)$ satisfy the commutation
relation
\begin{equation}
[\hat{X}_{1}(t),\hat{X}_{2}(t) ]=\frac{i}{2},
\end{equation}
which implies the Heisenberg uncertainty relation
\begin{equation}
\langle(\triangle \hat{X}_{1}(t))^{2}\rangle \langle(\triangle
\hat{X}_{2}(t))^{2}\rangle \geq \frac{1}{16}.
\end{equation}
A state of the radiation field is said to be squeezed whenever
\begin{equation}
\langle(\triangle \hat{X}_{i})^{2}\rangle<\frac{1}{4},(i=1 or 2),
\end{equation}
where
\begin{equation}
\langle(\triangle
\hat{X}_{i})^{2}\rangle=\langle\hat{X}_{i}^{2}\rangle-\langle\hat{X}_{i}\rangle^{2},
(i=1, 2).
\end{equation}
The degree of squeezing can be measured by the squeezing
parameter $S_{i}, (i=1, 2)$ defined by
\begin{equation}
S_{i}(t)=4\langle(\triangle \hat{X}_{i}(t))^{2}\rangle-1,
\end{equation}
which can be expressed in terms of the annihilation and creation
operators, $\hat{a}$ and $\hat{a}^{\dag}$ as follows
\begin{eqnarray}
S_{1}(t )= && (\langle \hat{a}^{2}(t ) \rangle-\langle \hat{a}(t
)\rangle^{2})\exp(2i\omega t)+(\langle \hat{a}^{\dagger2}(t )
\rangle-\langle \hat{a^{\dagger}}(t )\rangle^{2})\\
\nonumber  \times &&\exp(-2i\omega t) + 2( \langle \hat{ a}
^{\dagger}(t ) \hat{a}(t ) \rangle-\langle \hat{a} ^{\dagger}(t
)\rangle \langle\hat{a}(t ) \rangle),
\end{eqnarray}
and
\begin{eqnarray}
S_{2}(t )= && -(\langle \hat{a}^{2}(t ) \rangle-\langle \hat{a}(t
)\rangle^{2})\exp(2i\omega t)-(\langle
\hat{a}^{\dagger2}(\vec{p},t )
\rangle-\langle \hat{a^{\dagger}}(t )\rangle^{2})\\
\nonumber \times && \exp(-2i\omega t)+  2( \langle \hat{ a}
^{\dagger}(t ) \hat{a}(t ) \rangle-\langle \hat{a} ^{\dagger}(t
)\rangle \langle\hat{a}(t ) \rangle).
\end{eqnarray}
Then, the condition for squeezing in the quadrature components can
 be simply written as $S_{i}(t)<0$.
Now we obtain $\langle \hat{a}(t
)\rangle$, $\langle \hat{a} ^{\dagger}(t
)\rangle $ and $\langle \hat{ a}
^{\dagger}(t ) \hat{a}(t ) \rangle$
\begin{equation}
\langle \hat{a}(t
)\rangle=Tr_{field} ( \hat{\rho}_{field}\hat{a}(t
)),\langle \hat{a}^{\dagger} (t
)\rangle=Tr_{field} ( \hat{\rho}_{field}\hat{a}^{\dagger} (t
)),
\end{equation}
\begin{equation}
\langle \hat{a}^{\dagger} (t
)\hat{a}(t
)\rangle=Tr_{field} ( \hat{\rho}_{field}\hat{a}^{\dagger} (t
)\hat{a}(t
)),
\end{equation}
so that by using (75) we have
\begin{equation}
\langle \hat{a}(t
)\rangle=\int d^{3}p\sum_{k=0}^{\infty}\sum_{n=0}^{\infty} \frac{(2 \gamma t)^{k} }{k!}(\langle n | \hat{M}_{11} ^{k}(t)\hat{a}|n \rangle +\langle n | \hat{M}_{22} ^{k}(t)\hat{a}|n \rangle )=\langle \hat{a}^{\dagger} (t
)\rangle^{*} ,
\end{equation}
\begin{equation}
\langle \hat{a}(t
)^{2} \rangle=\int d^{3}p\sum_{k=0}^{\infty}\sum_{n=0}^{\infty} \frac{(2 \gamma t)^{k} }{k!}(\langle n | \hat{M}_{11} ^{k}(t)\hat{a}^{2} |n \rangle +\langle n | \hat{M}_{22} ^{k}(t)\hat{a}^{2} |n \rangle )=\langle \hat{a}^{\dagger 2} (t
)\rangle^{*},
\end{equation}
\begin{equation}
\langle \hat{a}^{\dagger} (t
)\hat{a}(t
)\rangle=\int d^{3}p\sum_{k=0}^{\infty}\sum_{n=0}^{\infty} \frac{(2 \gamma t)^{k} }{k!}n(\langle n | \hat{M}_{11} ^{k}(t) |n \rangle +\langle n | \hat{M}_{22} ^{k}(t) |n \rangle )
\end{equation}
where
\begin{eqnarray}
\langle n | \hat{M}_{11} ^{k}(t)\hat{a}|n \rangle &&= \sqrt{n} g_{+}^{k} (n,t)g_{+}^{k} (n-1,t)\psi_{1}(n,t) \psi_{1}^{*} (n-1,t)\\
\nonumber &&+\sqrt{n(n+1)} v_{+}^{'k} (n+1,t)g_{+}^{k} (n-1,t)\psi_{2}(n+1,t)\\
\nonumber && \times \psi_{1}^{*} (n-1,t) +nv_{+}^{'k} (n,t)g_{+}^{k} (n,t) \psi_{1}(n,t)\psi_{2}^{*} (n,t)\\
\nonumber &&  +n\sqrt{n+1} v_{-}^{'k} (n+1,t)v_{-}^{'k} (n,t)\psi_{2}(n+1,t) \psi_{2}^{*} (n,t),
\end{eqnarray}
\begin{eqnarray}
\langle n | \hat{M}_{11} ^{k}(t)\hat{a}^{2} |n \rangle &&= \sqrt{n(n-1)} g_{+}^{k} (n,t)g_{+}^{k} (n-2,t)\psi_{1}(n,t)\\
\nonumber && \psi_{1}^{*} (n-2,t)+\sqrt{n(n-1)(n+1)} v_{+}^{'k} (n+1,t)g_{+}^{k} (n-2,t)\\
\nonumber && \times \psi_{2}(n+1,t) \psi_{1}^{*} (n-2,t)+(n-1)\sqrt{n} v_{-}^{'k} (n-1,t)g_{+}^{k} (n,t)\\
\nonumber && \times \psi_{1}(n,t) \psi_{2}^{*} (n-1,t)+(n-1)\sqrt{n(n+1)} v_{-}^{'k} (n+1,t)\\
\nonumber && \times v_{-}^{'k} (n-1,t) \psi_{2}(n+1,t) \psi_{2}^{*} (n-1,t),
\end{eqnarray}
\begin{eqnarray}
\langle n | \hat{M}_{22} ^{k}(t)\hat{a}|n \rangle &&= \sqrt{n} g_{-}^{k} (n,t)g_{-}^{k} (n-1,t)\psi_{2}(n,t) \psi_{2}^{*} (n-1,t)\\
\nonumber &&+\sqrt{n(n-1)} v_{+}^{'k} (n-1,t)g_{+}^{k} (n,t)\psi_{2}(n,t)\\
\nonumber && \times \psi_{1}^{*} (n-2,t) +nv_{-}^{'k} (n,t)g_{-}^{k} (n-1,t) \psi_{1}(n-1,t)\psi_{2}^{*} (n-1,t)\\
\nonumber &&  +n\sqrt{n-1} v_{-}^{'k} (n-1,t)v_{-}^{'k} (n,t)\psi_{1}(n-1,t) \psi_{2}^{*} (n-2,t),
\end{eqnarray}
\begin{eqnarray}
\langle n | \hat{M}_{22} ^{k}(t)\hat{a}^{2} |n \rangle &&= \sqrt{n(n-1)} g_{-}^{k} (n,t)g_{-}^{k} (n-2,t)\psi_{2}(n,t) \psi_{2}^{*} (n-2,t)\\
\nonumber &&+\sqrt{n(n-1)(n-2)} v_{+}^{'k} (n-2,t)g_{-}^{k} (n,t)\psi_{2}(n,t)\\
\nonumber && \times  \psi_{1}^{*} (n-3,t)+n\sqrt{n-1} v_{-}^{'k} (n,t)g_{+}^{k} (n-2,t)\\
\nonumber && \times \psi_{1}(n-1,t) \psi_{2}^{*} (n-2,t)+(n-1)\sqrt{(n-1)(n-2)} v_{-}^{'k} (n,t)\\
\nonumber && \times v_{-}^{'k} (n-2,t) \psi_{1}(n-1,t) \psi_{1}^{*} (n-3,t),
\end{eqnarray}
 \hspace*{00.5 cm}In Fig.5 we have plotted the squeezing parameter $S_{1}(t)$ versus
the scaled time $\lambda t$ for three values of the parameter
$\vec{q}.\vec{g}$. As it is seen,  the quadrature component
$\hat{X}_{1}$ exhibits squeezing in the course of time evolution
when the influence of the gravitational field is negligible. With increase of the parameter $\vec{q}.\vec{g}$,
the parameter $S_{1}$ shows damped oscillatory behaviour and there is no quadrature squeezing.\\
\section{Summary and conclusions}
In this paper we studied the dissipative dynamics of the JCM with phase damping in the presence of a
classical homogeneous gravity field. The model consists of a moving two-level atom simultaneously
exposed to the gravitational field and a single-mode traveling radiation field in the presence of the phase damping.
 We presented a quantum
treatment of the internal and external dynamics of the atom based on an alternative su(2) dynamical algebraic
structure. By making use of the super-operator technique, we obtained an exact solution of the master equation for the
density operator of the quantized atom-radiation system, under the Markovian approximation.
Assuming that initially the
radiation field is prepared in a coherent state and the two-level atom is
in the excited state, we investigated the
influence of gravity on the temporal evolution of collapses and revivals of the atomic
population inversion, atomic dipole squeezing, atomic momentum
diffusion, photon counting statistics and quadrature squeezing of
the radiation field in the presence of the phase damping.
The results are summarized as follows: 1) the Rabi-like oscillations
in the atomic population inversion disappear, 2) the dipole squeezing decays
with increase of the parameter $\vec{q}.\vec{g}$, 3) in the
presence of the gravitational field, the atom can experiences
larger light-induced forces during its interaction with the
radiation field , 4) with increase of $\vec{q}.\vec{g}$, the
sub-Poissonian behaviour of the cavity-field is suppressed and it
exhibits super-Poissonian statistics and after some time, the Mandel parameter $Q(t)$ is stabilized at an asymptotic
zero value; the larger the parameter $\vec{q}.\vec{g}$ is more rapidly $Q(t)$ reaches the asymptotic value zero,
and 5) the quadrature
squeezing of the cavity-field disappears.\\
\\
{\bf  Acknowledgements} \\
One of the authors (M.M) wishes to thank The Office of Graduate
Studies of the Science and Research Campus Islamic Azad
University of Tehran for their support.

\vspace{20cm}
% ==================================================================================

{\bf FIGURE CAPTIONS:}

{\bf FIG. 1 } Time evolution of the atomic population inversion
versus the scaled time $\lambda t$. Here we have set
$q=10^{7}m^{-1}$,\\
$M=10^{-26}kg$,$g=9.8\frac{m}{s^{2}}$,$\omega_{rec}=.5\times10^{6}\frac{rad}{s}$,\\$\lambda=1\times
10^{6}\frac{rad}{s}$,
$\triangle_{0}=8.5\times10^{7}\frac{rad}{s}$,
 $ \varphi=0$, $\alpha=2$,
$\Delta=1.8\times 10^{6}\frac{rad}{s}$, $\gamma=7\times10^{-5}\frac{rad}{s}$,\\

 {\bf a)}For $\vec{q}.\vec{g}=0$.

{\bf b)}For $\vec{q}.\vec{g}=0.5 \times 10^{7}$.

{\bf c)}For $\vec{q}.\vec{g}=1.5 \times 10^{7}$.\\

{\bf FIG. 2 } Time evolution of the atomic dipole squeezing versus
the scaled time $\lambda t$ with the same corresponding data
 used in Fig.1;\\

{\bf a)}For $\vec{q}.\vec{g}=0$.

{\bf b)}For $\vec{q}.\vec{g}=0.5 \times 10^{7}$.

{\bf c)}For $\vec{q}.\vec{g}=1.5 \times 10^{7}$.\\

{\bf FIG. 3 } Time evolution of the atomic momentum diffusion
versus the scaled time $\lambda t$ with the same corresponding
data
 used in Fig.1;\\

{\bf a)}For $\vec{q}.\vec{g}=0$.

{\bf b)}For $\vec{q}.\vec{g}=0.5 \times 10^{7}$.

{\bf c)}For $\vec{q}.\vec{g}=1.5 \times 10^{7}$.\\

{\bf FIG. 4 } Time evolution of the Mandel parameter $Q(t)$ versus
the scaled time $\lambda t$ with the same corresponding data
 used in Fig.1;\\

{\bf a)}For $\vec{q}.\vec{g}=0$.

{\bf b)}For $\vec{q}.\vec{g}=0.5 \times 10^{7}$.

{\bf c)}For $\vec{q}.\vec{g}=1.5 \times 10^{7}$.\\

{\bf FIG. 5 } Time evolution of the squeezing parameter $S_{1}(t)$
versus the scaled time $\lambda t$ with the same corresponding
data
 used in Fig.1;\\

{\bf $a$)} For
$\vec{q}.\vec{g}=0$.

{\bf $b$)} For
$\vec{q}.\vec{g}=0.5 \times 10^{7}$.

{\bf $c$)} For
$\vec{q}.\vec{g}=1.5 \times 10^{7}$.
%\begin{figure}[]

%\epsfxsize=12cm\centerline{\epsffile{colors.eps}}
%\includegraphics[angle=0,scale=0.7]{p2-1a.eps}
 % \caption[]{fig-a1}\label{a1}
%\end{figure}

\end{document}